\documentclass[aps, prb,twocolumn,10pt,showpacs,superscriptaddress,longbibliography]{revtex4-2}
\usepackage{hyperref}
\hypersetup{
    colorlinks = false,
    pdfborder = 0 0 0
}
\usepackage{amsmath}
\usepackage{amsfonts}
\usepackage{amssymb}
\usepackage{pifont}
\usepackage{mathtools}
\usepackage{bm}
\usepackage{cases}
\usepackage{braket}
\usepackage[version=4]{mhchem}
\usepackage{graphicx}
\allowdisplaybreaks%

\begin{document}
\title{Emergent Halperin-Saslow mode and Gauge Glass \\
in quantum Ising magnet TmMgGaO$_4$}
 \author{Chun-Jiong Huang}
\affiliation{Department of Physics and HKU-UCAS Joint Institute 
for Theoretical and Computational Physics at Hong Kong, 
The University of Hong Kong, Hong Kong, China}
\author{Xiaoqun Wang}
\affiliation{School of Physics and Astronomy, Tsung-Dao Lee Institute, 
Shanghai Jiao Tong University, Shanghai 200240, China}
\affiliation{Key Laboratory of Artificial Structures and Quantum Control of MOE,
Shenyang National Laboratory for Materials Science, Shenyang 110016, China} 
\author{Ziqiang Wang}
\affiliation{Department of Physics, Boston College, Chestnut Hill, 
Massachusetts 02467, USA}
\author{Gang Chen}
\affiliation{Department of Physics and HKU-UCAS Joint Institute for Theoretical and 
Computational Physics at Hong Kong, The University of Hong Kong, Hong Kong, China}
\affiliation{State Key Laboratory of Surface Physics and Department of Physics, 
Institute of Nanoelectronics and Quantum Computing, 
Fudan University, Shanghai, 200433, China}
 
\date{\today}

\begin{abstract}
We propose quenched disorders could bring novel quantum excitations and models 
to certain quantum magnets. Motivated by the recent experiments on the quantum 
Ising magnet TmMgGaO$_4$, we explore the effects of the quenched disorder and 
the interlayer coupling in this triangular lattice Ising antiferromagnet. It is pointed 
out that the weak quenched (non-magnetic) disorder would convert the emergent 
2D Berezinskii–Kosterlitz–Thouless (BKT) phase and the critical region into a gauge 
glass. There will be an emergent Halperin-Saslow mode associated with this gauge 
glass. Using the Imry-Ma argument, we further explain the fate of the finite-field 
$C_3$ symmetry breaking transition at the low temperatures. The ferromagnetic
interlayer coupling would suppress the BKT phase and generate a tiny ferromagnetism. 
With the quenched disorders, this interlayer coupling changes the 2D gauge glass 
into a 3D gauge glass, and the Halperin-Saslow mode persists. This work merely 
focuses on addressing a phase regime in terms of emergent U(1) gauge glass behaviors 
and hope to inspire future works and thoughts in weakly disordered frustrated magnets 
in general. 
\end{abstract}

\maketitle

Disorder is unavoidable in quantum materials. Topological phases, such 
as the gapped spin liquids with intrinsic topological orders and topological insulators
with symmetry-protected topological properties, are robust against weak
disorders, so disorder effects are not often invoked in these 
contexts~\cite{Wen_2013,RevModPhys.82.3045}. For conventional ordered phases, 
critical phases, and phase transitions, 
disorder effects require a more serious consideration. This is  
particularly relevant for many frustrated materials. Clearly,
disorder and frustration play an important role in spin glass 
that is unambiguously one of most challenging subjects in condensed matter physics~\cite{RevModPhys.58.801}. 
Quenched impurities were shown to select spiral magnetism among 
degenerate spiral states in classically frustrated magnets~\cite{PhysRevB.84.064438}, 
and quenched disorders could bring a strong quantum entanglement and 
generate highly entangled quantum states~\cite{PhysRevLett.118.087203}. 
However, there have not yet been many 
solid and conclusive results for the effects of quenched/annealed disorders 
in frustrated quantum magnetism~\cite{Henley_2001}. 
An impressive example of the disorder effect is the spin-1/2 random Heisenberg chain 
that was studied with the real-space renormalization group analysis and 
the master equation by D.S. Fisher~\cite{PhysRevB.50.3799}.
It establishes that any weak randomness in the exchange due to disorders 
would drive the system into the strong disordered random fixed point. 
More substantially, this analysis is asymptotically exact. 
For disorder effects, one should distinguish the long-distance and 
low-energy physics from the short-distance and high-energy one. 
The numerical study of small systems with buried disorders 
is certainly difficult to provide much useful information about the former
with D.S. Fisher's remarkable result as an example, while the understanding of the 
latter is feasible. Thus, theoretical arguments or making connection to established 
results in statistical physics can sometimes be helpful.

\begin{figure}[t]
\includegraphics[width=0.9\linewidth]{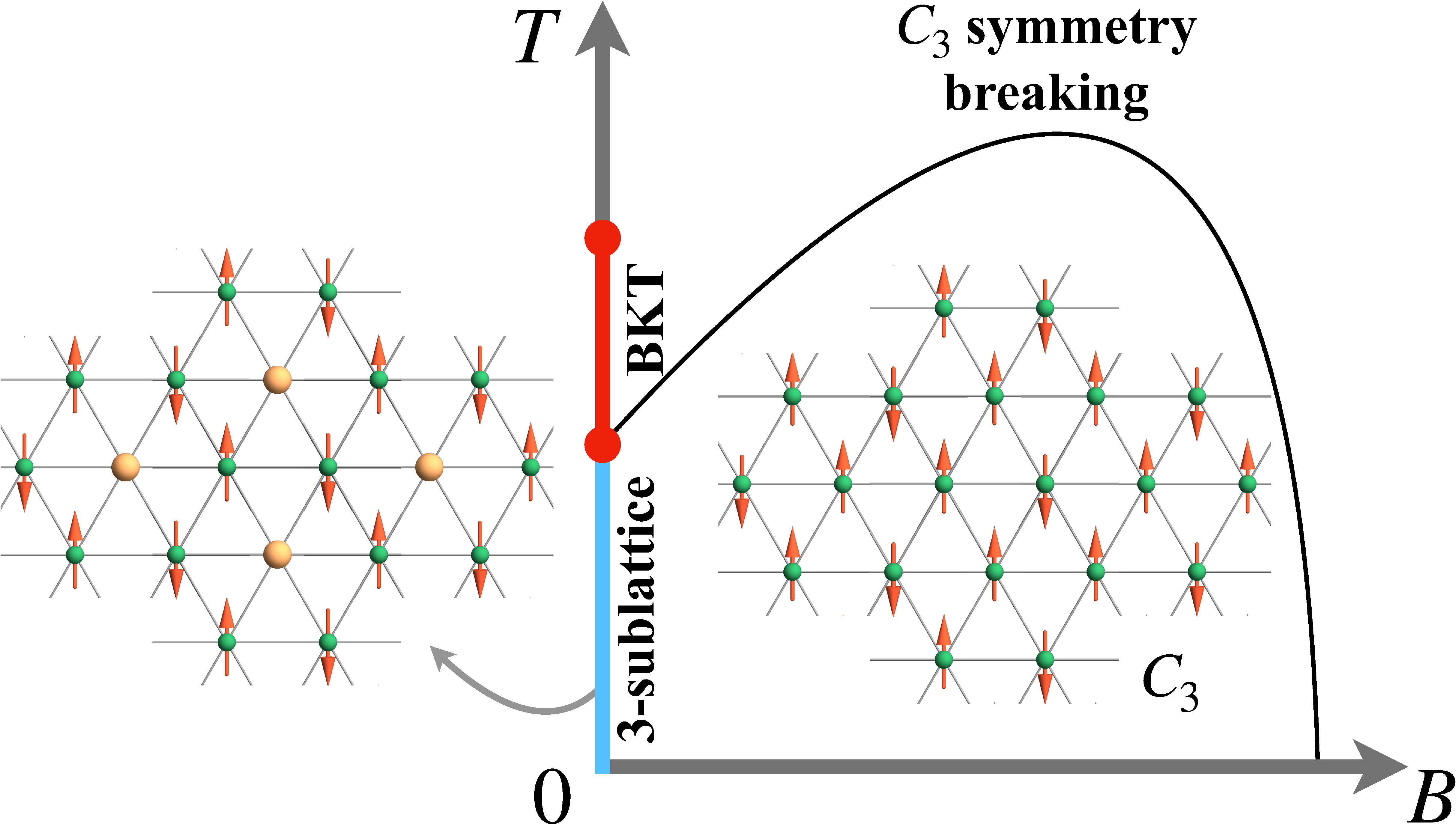}
\caption{
The schematic phase diagram of the TFIM in the magnetic field and the temperature. 
The inset are the $\langle S^z \rangle$ configuration in the three-sublattice order 
at low temperatures and the ``up-up-down'' state in the intermediate magnetic field. 
The orange ball refers to the zero $\langle S^z \rangle$. In both plots, the transverse 
component is polarized by the intrinsic magnetic field. The solid line is a strongly 
first-order transition that breaks the $C_3$ rotation. 
} 
\label{fig1}
\end{figure}

We are partly motivated from the triangular lattice antiferromagnet 
TmMgGaO$_4$~\cite{Cevallos2018,Shen2019intertwined,Li2018absence,Li2020partial,
Li2019KTmelting,Liu2020intrinsic,Hu2020evidence,Dun2020neutron}.
TmMgGaO$_4$ is a Mott insulator where the magnetic Tm$^{3+}$ ions 
form a triangular lattice. This material is isostructural to the triangular lattice 
spin liquid candidate YbMgGaO$_4$ that caught some attention earlier~\cite{Li2015rareearth,Li_2015,Shen2016evidence,Paddison2017continuous,Li2016MSR,
Xu2016absence,PhysRevB.94.035107,PhysRevB.96.054445,PhysRevB.96.075105,
PhysRevB.97.125105,Li2017crystalline,Zhu2017disorder,Shen2018fractionalized,
Zhang2018hierarchy,PhysRevX.8.031028}. TmMgGaO$_4$ was shown to have an
anisotropic magnetic behavior with a nearly zero magnetic response to 
the external magnetic field in the triangular plane~\cite{Cevallos2018} and
a large magnetic moment normal to the triangular planes, {\sl i.e.} the $z$ direction. 
Despite the structural disorders due to the Tm$^{3+}$ positional disorder 
and the Ga-Mg site disorder~\cite{Cevallos2018}, further measurements 
found an ordering transition  $\sim$1K~\cite{Shen2019intertwined,Li2020partial,Dun2020neutron}. 
Detailed neutron scattering revealed a dominant 
Bragg peak at the $K$ point of the Brillouin zone, suggesting a 
three-sublattice magnetic order. Our previous theoretical efforts proposed 
a weakly-split doublet for the Tm$^{3+}$ ion to understand 
the low-temperature magnetic properties~\cite{Shen2019intertwined,PhysRevResearch.1.033141}, 
instead of the conventional non-Kramers doublet~\cite{Liu2020intrinsic,PhysRevB.98.045119,Li2018absence}. 
The weak splitting between the doublet was modeled as 
an intrinsic transverse field on the effective spin-1/2 local 
moment ${\boldsymbol S}_i$, and the exchange is primarily Ising-like
with~\cite{Shen2019intertwined,PhysRevB.98.045119} 
\begin{equation}
H_{\text{TFIM}} = \sum_{ij } J_{ij} S^z_i S^z_j - h\sum_i S^y_i.
\end{equation}
This transverse field Ising model (TFIM) with an {\sl intrinsic microscopic origin} 
provides a reasonable explanation of both the magnetic structure and 
the magnetic excitation from the inelastic neutron 
scattering measurements~\cite{Shen2019intertwined}. The intrinsic transverse 
field in $H_{\text{TFIM}}$ generates quantum tunneling events between 
different classical Ising ground states in Fig.~\ref{fig1}, and the resulting 
three-sublattice order is a quantum effect, known as the 
``order-by-disorder''~\cite{Moessner2000,Moessner2001ising,Isakov2003interplay}. 
Due to the multipolar nature of the local moments, only the $z$ component, 
$S^z$, is visible in the magnetic measurement~\cite{Shen2019intertwined,Liu2020intrinsic,Li2018absence}. 
It is understood that $S^z$ flips the transverse component and generates 
quantum dynamics for the inelastic neutron detection~\cite{Shen2019intertwined}. 
More quantitative aspects of the experiments can be more sensitive 
to disorder effects on the exchange and Land\'e factors as well as the 
residual coupling to the high-order multipolar moments~\cite{Li2020partial,Dun2020neutron,Cevallos2018}.

Aligned with the well-known result of the triangular lattice 
TFIM~\cite{Moessner2001ising,Isakov2003interplay}, 
a finite-temperature Berezinskii–Kosterlitz–Thouless (BKT) phase 
with an emergent U(1) symmetry would exist above the low-temperature 
three-sublattice order~\cite{Isakov2003interplay,Liu2020intrinsic,Li2019KTmelting}. 
In an interesting Ga-based NMR experiments on TmMgGaO$_4$~\cite{Hu2020evidence}, 
a hump in the $1/T_1$ dynamics was found from ${\sim}$0.9K to ${\sim}$1.9K, 
indicating significant low-energy spin excitations in this temperature window, 
and was attributed as an evidence for the BKT phase, while there is
no such hump in the Knight shift.  
As the magnetic susceptibility in the BKT phase exhibits a power-law 
divergence~\cite{PhysRevLett.115.127204} with the field in the zero field limit, 
Ref.~\cite{Hu2020evidence} 
then studied the magnetic susceptibility from 0.6T-0.9T and 
numerically fitted to power laws. This field range, 
equivalent to $\sim$4K-$6$K
for the Tm moment, may be a bit too large to be perturbative and
would drive the system into a ``up-up-down'' state in Fig.~\ref{fig1}~\cite{Liu2020intrinsic}.  
Ref.~\cite{Dun2020neutron} performed an advanced neutron 
scattering measurement on TmMgGaO$_4$ and extracted 
the spin correlation length, $\xi$ 
as a function of temperature. They found that, $\xi$ is $\sim$3-7 lattice 
spacing in the claimed BKT phase, and saturates to $\sim$30 lattice spacing 
in the zero-temperature limit. This finite $\xi$ is in contrary to the divergent 
$\xi$ expected for the BKT phase. 

\begin{figure}[t]
\includegraphics[width=0.85\linewidth]{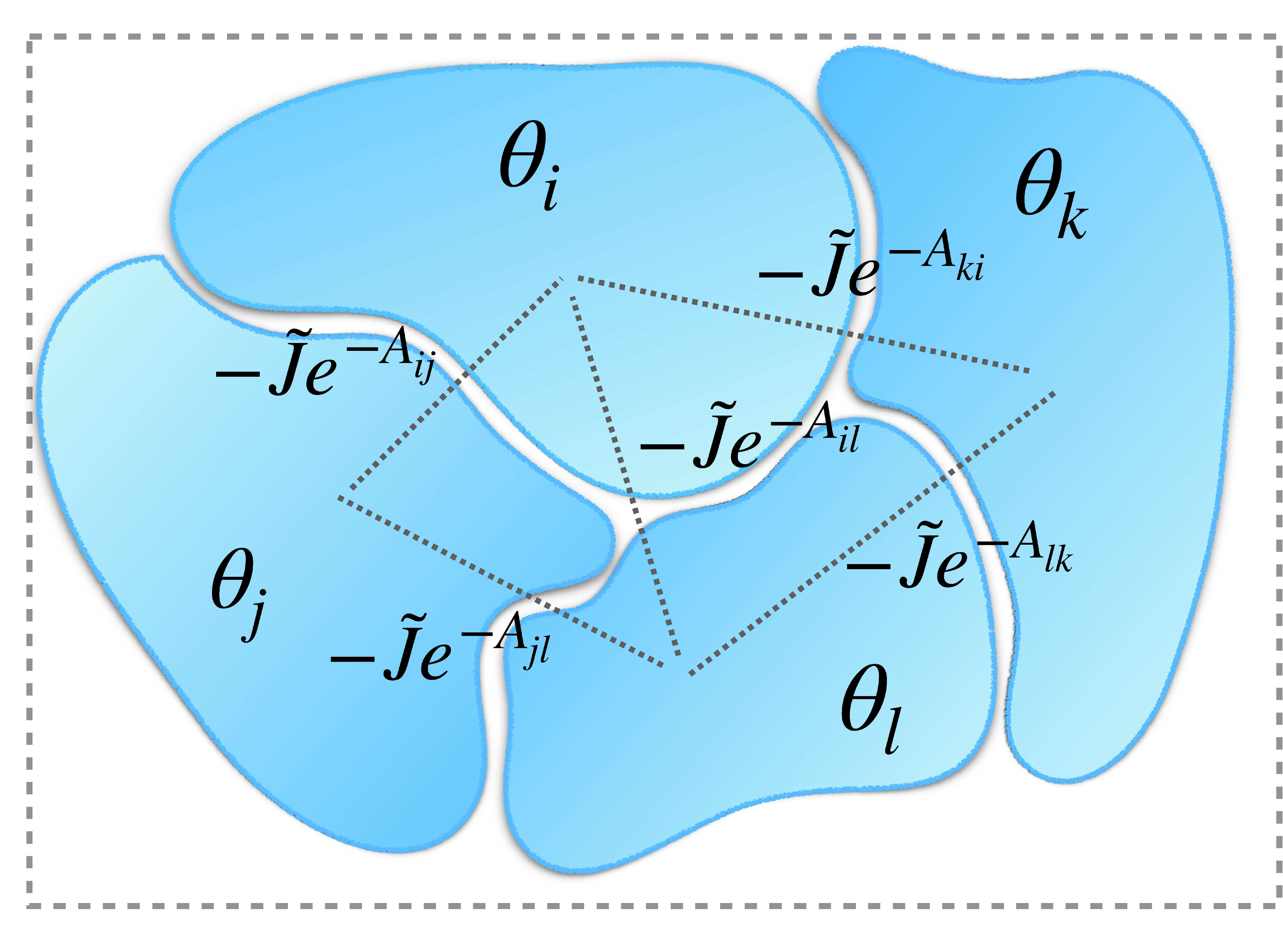}
\caption{The schematic plot of phase puddles and couplings for the U(1) gauge glass. } 
\label{fig2}
\end{figure}

Our work is partly to resolve different pieces of experiments in TmMgGaO$_4$. 
Considering several suggestion of non-magnetic disorders~\cite{Dun2020neutron,Li2020partial,Cevallos2018},
we begin with the analysis of the zero field regime for the single triangular layer and 
incorporate the non-magnetic disorder by assuming a weak disorder.
This is a continuation of the disorder analysis that was advertised 
in the end of Ref.~\onlinecite{Liu2020intrinsic}. We further include 
the interlayer coupling and consider the three dimensional limit. 
As a convention, we assume the disorders are uncorrelated and local. 
In the clean limit, the low-energy physics of the triangular lattice TFIM is 
well captured by a coarse-grained model~\cite{PhysRevLett.115.127204} 
\begin{eqnarray}
H \simeq - J \sum_{\langle ij \rangle} \cos (\theta_i -\theta_j) - J_6 \sum_i \cos (6 \, \theta_i ),
\label{eq1}
\end{eqnarray}
where $\langle ij \rangle$ refers to the nearest-neighbor bond on the 
coarse-grained triangular lattice, and $\theta$ is the phase of the complex 
order parameter at the $K$ point.  Here, the order parameter is given as the 
  three-sublattice order parameter ${\psi = m_1 + m_2 e^{i 2\pi/3 } + m_3 e^{-i 2\pi/3}}$,
  where $m_i$ (${i=1,2,3}$) refers to $\langle  S^z \rangle$ of three sublattices at 
  the neighboring sites, and ${\theta= arg[\psi]}$, where we have set the lattice constant 
  to unity. The coupling $J$, ${J>0}$, is the coarse-grained exchange 
  and favors a uniform phase $\theta$.
  The second term is a dangerously irrelevant 6-fold anisotropic term~\cite{Jose1977renormalization,Blankschtein1984ordering,Caselle1998stability,
  Caselle1998stability,Lou2007emergence}. 
  It is irrelevant in the BKT phase, and there is an emergent U(1) symmetry. 
  At the low temperatures below the BKT phase, it pins the system to the three-sublattice order. 
  We expect ${J_6 <0}$ for TmMgGaO$_4$. In the BKT phase, if the spacing
  in the coarse-grained lattice is large enough, $J_6$ becomes very small and 
  can be neglected in the initial analysis. As the BKT phase is compressible,   
  it is more susceptible to the disorder.   
  Once the non-magnetic disorder is considered, the coarse-grained model
  for the phase $\theta$ takes the following effective form, 
  \begin{eqnarray}
H_{gg} \simeq - \tilde{J} \sum_{\langle ij \rangle} \cos (\theta_i -\theta_j - A_{ij})  ,
\label{eq2}
\end{eqnarray}
where $i,j$ here should be interpreted as different phase puddles due to disorder for 
the U(1) phase $\theta$ (see Fig.~\ref{fig2}), and the gauge field $A_{ij}$ takes care 
of the disorder effects and is a random variable from 0 to $2\pi$. 
The parameter $\tilde{J}$, with ${\tilde{J}>0}$,  is an effective coupling between the puddles, 
and the neglecting of the randomness in the magnitude does not qualitatively change the physics. 
The puddle size is determined by the disorder strength. The model in Eq.~\eqref{eq2} is 
known as the Hamiltonian of the U(1) gauge glass model in two 
dimensions~\cite{Daniel1988,Ebner1985diamagnetic,Fisher1991vortex}. 
It should be a good approximation of the physics as long as the anisotropic term 
becomes negligibly small at the scale of the puddle size, and we do not really need to sit right on 
the BKT phase in the clean limit. Numerical studies have shown that 
the lower critical dimension of this model is greater than 2, 
but less than 3~\cite{PhysRevB.69.184512,PhysRevB.66.224507}. 
The model in Eq.~\eqref{eq2} without the dangerously 
irrelevant anisotropic term would not have a thermal transition in the $2d$ limit. 
At low temperatures, the anisotropic term is expected to pin $\theta$ 
and generate the three-sublattice order.

The well-known excitation of the U(1) gauge glass is the gapless Halperin-Saslow 
mode~\cite{PhysRevB.16.2154,PhysRevB.79.214436}. 
This mode is related to the slow and smooth variation of the phase angle $\theta$ and is of 
the Goldstone-type spin wave mode. It has been proposed in the context of the triangular 
lattice antiferromagnets NiGa$_2$S$_4$ and FeAl$_2$Se$_4$ with consistent  
thermodynamic behaviors~\cite{PhysRevB.79.140402,Nakatsuji1697,PhysRevB.99.054421,
PhysRevB.79.214436,PhysRevLett.105.037402}. 
Over there, the system is expected to have a continuous 
spin rotational symmetry and the exchange scale is fairly large so that    
the low-energy properties of this mode can be well-accessed~\cite{PhysRevB.79.140402,PhysRevB.99.054421}. 
In our case, the U(1) symmetry is emergent and is not microscopically originated,
thus the Halperin-Saslow mode is emergent and will in principle develop a tiny gap 
due to the irrelevant anisotropic term in Eq.~\eqref{eq1}. This tiny gap is determined by 
the magnitude of the anisotropic term at the puddle scale, and may not 
be well-resolved with the experimental resolution. Since this nearly-gapless emergent 
Halperin-Saslow mode is a spin-wave mode in nature, it shows up around the $K$ point 
in the inelastic neutron scattering measurement and contributes to the NMR $1/T_1$ dynamics.

The emergent U(1) symmetry is immediately destroyed once the magnetic 
field is applied along the $z$ direction. The system quickly breaks the 
Ising-triangle degeneracy and enters  an ``up-up-down'' state in the 
intermediate magnetic field regime~\cite{Liu2020intrinsic} (see Fig.~\ref{fig1}).  
The system no longer has the time reversal symmetry, and 
this state breaks the $C_3$ rotational symmetry around the centers 
of the triangular plaquettes. The Ginzburg-Landau 
analysis yields a strongly first-order transition, and this result was  
supported by the numerical calculation~\cite{Liu2020intrinsic}. 
As the frustration is lifted in this finite field regime, 
the transition temperature is enhanced compared to the 
zero-field one. From the symmetry point of view, 
the non-magnetic disorder would directly couple to the $C_3$ order parameters, 
e.g. the nearest-neighbor bond spin correlation.
Therefore, these non-magnetic disorders behave like random fields  
for the $C_3$ symmetry breaking. 
If we restrict ourselves to a single Tm$^{3+}$ layer and neglect the weak interlayer coupling, 
the standard Imry-Ma argument~\cite{PhysRevLett.35.1399} suggests that the long-range $C_3$ 
order should be destroyed by any arbitrarily weak disorder strength. The system would 
be separated into different domains of the $C_3$ orders, and the domain size is determined 
by the balance between the surface energy of the domains and the random field strength. 
The strong first order transition is expected to be rounded by the disorder. As
two dimensions is the lower critical dimension for the stability of the ordered phase 
based on the Imry-Ma's argument, the correlation length for the $C_3$ order
is an exponential function of the disorder strength below the transition temperature. 
This may be further measured by advanced neutron scattering techniques 
in Ref.~\onlinecite{Dun2020neutron}.

\begin{figure}[t]
\includegraphics[width=1.0\linewidth]{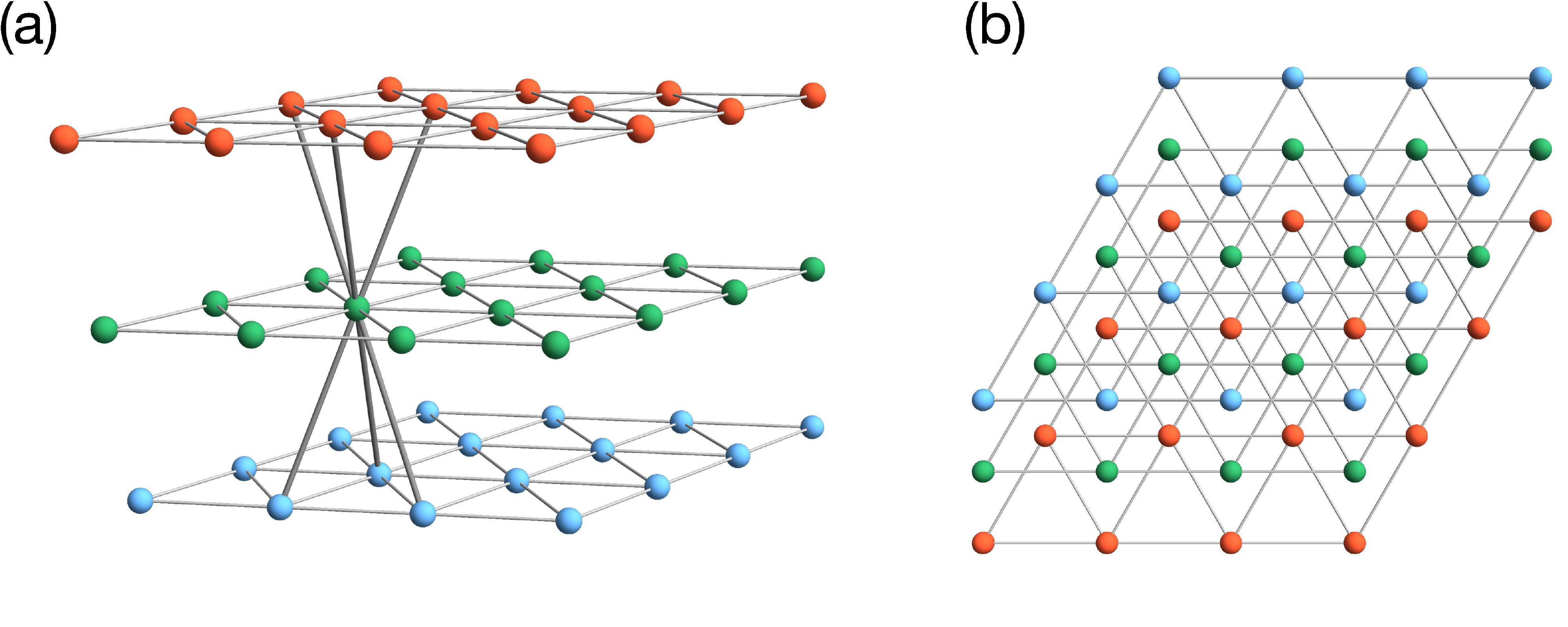}
\caption{(Color online.) 
(a) The ABC stacking of the multi-layered triangular lattice for the Tm$^{3+}$ ions. 
Each Tm$^{3+}$ moment is coupled to six nearest-neighbor Tm moments via the 
interlayer coupling. (b) The top view of the ABC stacking. 
} 
\label{fig3}
\end{figure}

The Tm$^{3+}$ local moment has a rather large magnetic moment and is  
approximately $7.3\mu_{\text B}$ per Tm$^{3+}$ ion~\cite{Cevallos2018}, 
suggesting the long-range dipole-dipole interaction can be particularly 
relevant for the low-temperature magnetic properties. 
Even for the clean limit in two dimensions, since the BKT phase 
has an infinite correlation length that suppresses the irrelevant
coupling, how the (infinitely) long-range dipole-dipole interaction 
impacts on the BKT phase is an interesting question. It was estimated 
that the second (third) neighbor dipole-dipole interaction 
in the triangular plane is $\sim 0.48$K (${\sim 0.31}$K)~\cite{Liu2020intrinsic}.  
It turns out the interlayer dipole-dipole interaction can be 
equally relevant as the orientation for the nearest-neighbor Tm bond
between the neighbor layers enhances the interaction and generates 
a ferromagnetic coupling. It is estimated that the nearest-neighbor interlayer 
coupling is $\sim -0.30$K, and there are six nearest-neighbor interlayer sites
(see Fig.~\ref{fig3}). This coupling is expressed 
as
\begin{eqnarray}
H_{\text{int}} = \sum_{\langle ij \rangle_{\text{int}}} J_{\text{int}}^{} S_i^z S_j^z ,
\end{eqnarray}
where ${J_{\text{int}} <0}$ and ``int'' refers to the interlayer coupling. Although
this coupling is relatively small compared to the dominant coupling for the intralayer 
TFIM, it is still natural to consider the role of this coupling,
especially for the very low temperature properties around $1$K. 
Once this interlayer coupling is considered, the system immediately becomes 
three dimensions. The BKT phase and transitions are defined in two dimensions
and will disappear in three dimensions (except the vortices form strong line objects), 
and then the three-dimensional ordering transition is often obtained from $T \sim \xi (T) ^2 |J_{\text{int}}| $
where $\xi (T)$ is the intralayer correlation length.  On the other hand, the 
ferromagnetic interlayer coupling for a three-dimensional system would favor 
ferromagnetism. For this purpose, we start from the BKT phase at the finite 
temperature for each layer and couple the layers with the interlayer coupling. 
We can obtain the coarse-grained free energy in this regime as 
\begin{eqnarray}
F_{3d}  &=& \sum_l \frac{M_l^2}{2\chi_{2d}} 
           - 3|J_{\text{int}}| \sum_{\langle ll' \rangle} M_l^{} M_{l'}^{} + c_4 \sum_l M_l^4
           \nonumber \\
&& + \cdots,
\label{eqfree}
\end{eqnarray}
where $M_l$ is the uniform magnetization per site on the $l$-th layer, and 
``$\cdots$'' refers to other terms such as the ones from the intralayer Ising 
coupling and the dipole-dipole interactions. The last term in $F_{3d}$ is 
simply the leading order term in the expansion of the free energy in $J_{\text{int}}$. 
This can also be interpreted as a mean-field treatment of the ferromagnetic 
interlayer coupling. This is similar to the ``chain mean-field theory''~\cite{Starykh_2015},
that has been successfully and widely applied to understand 
the experiments in low-dimensional magnetic systems~\cite{PhysRevLett.98.077205,PhysRevB.78.174420,PhysRevB.94.035154}, 
and is shown to be rather reliable. In the BKT regime of each layer, 
the uniform susceptibility $\chi_{2d}$ was argued to diverge~\cite{PhysRevLett.115.127204}, 
and the first term that describes the energy cost for a finite magnetization simply 
vanishes. Thus, a tiny ferromagnetism would be favored in addition 
to the three-sublattice antiferromagnetism once the weak ferromagnetic 
interlayer coupling is turned on in this BKT regime. In fact, a simple 
calculation of Eq.~\eqref{eqfree} would favor a weak ferromagnetism 
if ${\chi_{2d} > 1/(6|J_{\text{int}}| )}$. 
Thus, even if we relax the divergent $\chi_{2d}$ condition, 
a weak ferromagnetism should be expected. This 
above argument holds even with weak disorders.
We attribute this as the origin of the weak $\Gamma$ point 
Bragg peak in the neutron scattering measurement~\cite{Shen2019intertwined}. 

Once the interlayer coupling is included, the coarse-grained model in 
Eq.~\eqref{eq2} picks up an interlayer Josephson-type coupling,
and the model becomes a U(1) gauge glass model in three dimensions. 
Again, the interlayer Josephson-type coupling is also disordered by the 
Ga-Mg disorder and the Tm positional disorder. The emergent 
Halperin-Saslow mode persists to three dimensions.  It was 
numerically shown that this U(1) gauge glass model  
has a finite-temperature phase transition~\cite{PhysRevB.61.12467}. 
The ordering would be further solidified by the anisotropic term in Eq.~\eqref{eq1}.  

\emph{Discussion.}---Here we give a discussion about the 
specific properties of TmMgGaO$_4$, 
and then sketch a discussion about the disorder treatments in 
frustrated magnets and scenarios in spin liquids. As TmMgGaO$_4$ 
is cooled from the high-temperature paramagnetic regime, it gradually 
builds up spin correlations. According to the careful and detailed
measurements in Ref.~\onlinecite{Dun2020neutron}, the spin correlation length 
grows up to the Tm-Tm lattice spacing around 6K, and this can be regarded as 
the entering of the cooperative paramagnetic regime. When the correlation length grows 
up to the puddle size set by the disorder strength in Eq.~\eqref{eq2}, 
an approximate U(1) gauge glass description applies. In addition to the emergent 
Halperin-Saslow mode, there exists more subtle low-energy excitations, 
{\sl i.e.} ``droplets'' of the droplet theory for the gauge glass~\cite{Daniel1988,PhysRevB.43.130,PhysRevB.66.224507}. 
These are non-smooth deformation of the phase, and related to the vortex configurations 
that re-arrange, and when they do on length scale $L$ it costs an energy that depends 
 on $L$ algebraically and on the dimensions. These low-energy excitations all contribute 
 to the specific heat, 
while the Halperin-Saslow mode should be detectable in the inelastic neutron scattering and 
NMR $1/T_1$ measurements. Thus,
although the Ga-Mg disorder may cause a further complication in deciphering the experiment,
 we propose the hump in the NMR $1/T_1$ dynamics~\cite{Hu2020evidence}
is a signature of the emergent Halperin-Saslow mode for the gauge glass. 
Compared to the BKT scenario in the clean two-dimensional case~\cite{Hu2020evidence}, 
the Halperin-Saslow mode of gauge glass is out of the weakly-disordered perspective 
and holds even for three dimensions. As the Halperin-Saslow mode
would scatter the phonons strongly, 
we expect there will be a dip/valley-like suppression in the thermal transport. 
Since the gap in the three-sublattice order is quite small~\cite{Shen2019intertwined}, 
this suppression may not be quite visible.  
As the system is further cooled, the dangerously irrelevant anisotropic term 
in Eq.~\eqref{eq1} becomes important and drives the system into the three-sublattice 
order at low temperatures. The low-temperature three-sublattice order in
TmMgGaO$_4$ is a fully gapped incompressible state and thus a bit more 
stable to weak disorders than compressible ones. Due to the quantum order
by disorder, the spin-wave gap of this order is small but is still
determined by the combination of the microscopic couplings. This 
spin-wave gap shows up directly in the inelastic neutron 
scattering~\cite{Shen2019intertwined,Dun2020neutron}, 
and as an Arrhenius form in the specific 
heat and the NMR $1/T_1$ dynamics at the corresponding 
temperatures~\cite{Hu2020evidence}. 
 
The study of disorder effects in frustrated magnets has not been very fruitful
as we are more familiar with periodic lattice models and continuum models
and neither remain with disorders. Due to the diversity of frustrated 
materials and the physical properties, the standard methods like Harris 
criteria, Imry-Ma argument, replica theory, real-space renormalization group
should be properly adjusted to different contexts and disorder realization. 
Thus, experiments are necessary to give the guidance for the theoretical development here.
This is especially so for the disorder effect in spin liquids which is an important subject along
the line. While weak disorders do not cause much effect for topological spin liquids, 
it can be important for the physical properties of critical spin liquids. 
The perturbative analysis of weak disorders for the spinon Fermi 
surface U(1) spin liquid indicates that the system changes from the spinon Fermi surface 
metal to the diffusive spinon metal and then to the spinon Anderson insulator as the disorder 
strength increases, while the fractionalization and emergent non-locality that define 
the spin liquids are actually preserved~\cite{PhysRevLett.109.077205,PhysRevB.96.054445}. 
In contrast, the strong disorder argument would 
directly favor a high dimensional version of the random singlet phase~\cite{Kimchi_2018,PhysRevX.8.031028}. 
Depending on the sample quality and disorder strengths, both regimes may be 
realized in relevant materials such as the doped semiconductors~\cite{PhysRevLett.109.077205} and
the triangular lattice antiferromagnet YbMgGaO$_4$~\cite{Li2015rareearth,Li_2015,Shen2016evidence,Paddison2017continuous,Li2016MSR,Xu2016absence,
PhysRevB.94.035107,PhysRevB.96.054445,PhysRevB.96.075105,
PhysRevB.97.125105,Li2017crystalline,Zhu2017disorder,Shen2018fractionalized,Zhang2018hierarchy}.

% interlayer coupling, C3 axes

\emph{Acknowledgments.}---We thank Yao Shen for informing the interlayer 
distance and angles, Rong Yu for a recent conversation, and Yong Baek Kim,   
Leon Balents, Matthew Fisher for the email communication. This work is 
supported by the Ministry of Science and Technology of China 
with Grant No. 2016YFA0300500, 2018YFE0103200, 
2016YFA0301001, by Shanghai Municipal Science and 
Technology Major Project with Grant No.2019SHZDZX04, 
and by the Research Grants Council of Hong Kong with 
General Research Fund Grant No.17303819 and No.17306520. 
Z.W. is supported by the U.S. Department of Energy, 
Basic Energy Sciences Grant No. DE-FG02-99ER45747.

\bibliography{refs}

\end{document}